\documentstyle[12pt]{article}
\begin{document}

\def\beqra{\begin{eqnarray}} \def\eeqra{\end{eqnarray}}
\def\beqast{\begin{eqnarray*}} \def\eeqast{\end{eqnarray*}}
\def\beq{\begin{equation}}	\def\eeq{\end{equation}}
\def\be{\begin{enumerate}}   \def\ee{\end{enumerate}}

\def\fnote#1#2{\begingroup\def\thefootnote{#1}\footnote{#2}\addtocounter
{footnote}{-1}\endgroup}

\def\ut#1#2{\hfill{UTTG-{#1}-{#2}}}

\def\sppt{Research supported in part by the
Robert A. Welch Foundation and NSF Grant PHY 9009850}

\def\utgp{\it Theory Group,  Department of Physics \\ University of Texas,
 Austin, Texas 78712}

\def\gam{\gamma}
\def\Gam{\Gamma}
\def\la{\lambda}
\def\eps{\epsilon}
\def\La{\Lambda}
\def\si{\sigma}
\def\Si{\Sigma}
\def\al{\alpha}
\def\Tha{\Theta}
\def\tha{\theta}
\def\vphi{\varphi}
\def\del{\delta}
\def\Del{\Delta}
\def\ab{\alpha\beta}
\def\om{\omega}
\def\Om{\Omega}
\def\mn{\mu\nu}
\def\mun{^{\mu}{}_{\nu}}
\def\kap{\kappa}
\def\rsi{\rho\sigma}
\def\beal{\beta\alpha}
\def\cO{{\cal O}}
\def\ltsim{\matrix{<\cr\noalign{\vskip-7pt}\sim\cr}}
\def\gtsim{\matrix{>\cr\noalign{\vskip-7pt}\sim\cr}}
\def\haf{\frac{1}{2}}
\def\lag{\langle}
\def\rag{\rangle}
\def\til{\tilde}
\def\rta{\rightarrow}
\def\eqv{\equiv}
\def\nab{\nabla}
\def\pa{\partial}
\def\npb#1#2#3{{\it Nucl. Phys.} {\bf B#1} (19#2) #3 }
\def\plb#1#2#3{{\it Phys. Lett.} {\bf B#1} (19#2) #3 }
\def\prc#1#2#3{{\it Phys. Rev.} {\bf C#1} (19#2) #3 }
\def\prd#1#2#3{{\it Phys. Rev.} {\bf D#1} (19#2) #3 }
\def\pr#1#2#3{{\it Phys. Rev.} {\bf #1} (19#2) #3 }
\def\prep#1#2#3{{\it Phys. Rep.} {\bf C#1} (19#2) #3 }
\def\prl#1#2#3{{\it Phys. Rev. Lett.} {\bf #1}(19#2) #3 }
\def\ul{\underline}
\def\indt{\parindent2.5em}
\def\nd{\noindent}

\vspace*{24pt}
\parindent=2.5em

\ut{04}{94}

\vspace{6pt}
\hfill\today

\begin{center}

\large{\bf GRAVITINOS AND A LOW ULTIMATE

\vspace{12pt}
TEMPERATURE FOR THE EARLY UNIVERSE}

\vspace{24pt}
 Willy Fischler\fnote{*}{\sppt}

\vspace{18pt}
 \utgp

\vspace{18pt}

\end{center}

\abstract{We obtain new stronger bounds by orders of magnitude, on the ultimate
temperature of the universe by exploiting the copious production of gravitinos
at finite temperature.}

\vspace{1.6in}
\begin{minipage}{5in}
\begin{center}
I dedicate this work to the memory of my grandparents \\{\it
Anna and Wolf Fischler} who perished in Auschwitz during the Holocaust.
\end{center}
\end{minipage}
\vfill

\pagebreak
\baselineskip=21pt
\setcounter{page}{1}

In this paper we will derive new bounds on the temperature of the universe by
considering the cosmological implication of gravitinos with masses varying from
10$^{-4}eV$ to 1TeV.  The difference with previous results [1-6, 10] is due to
the
large interactions of the spin 1/2 component of the gravitinos with the hot
primordial plasma.

There are two major assumptions in this paper:

\be
\item We will assume that the theory describing ultramicroscopic physics leaves
over at lower energies an $N=1$ supergravity theory [7].

\item We will also assume that the typical splitting in supermultiplets after
supersymmetry breaking is of $\cO\;(G_F^{-1/2})$.  This represents the
prejudice that supersymmetery ``solves'' the old fine tuning puzzle.
\ee

The low  energy spectrum then contains the usual particles and their
superpartners as well as a gravitino.  We will begin by discussing briefly the
different masses for superpartners of usual particles.  The origin for the mass
of the gravitino is the superhiggs mechanism [8] which is of gravitational
strength.  In contrast,  the masses of the other superpartners depend on how
supersymmetry breaking is fed down to the low energy sector and may not involve
to leading order the Planck mass,  $M_p$.  This is why the gravitino mass
$m_{3/2}$ can be quite different than the typical superpartner mass of
$\cO(G_F^{-1/2})$ if the physics that mediate supersymmetry breaking to the
usual
sector, is not gravitational.  Let me briefly review how masses of scalar
partners
and gauginos arise when gravitational feeddown  is negligible.  The masses of
gauginos come from dimension five operators of the form
\beq
\frac{1}{M}\,\int\,d^2 \Tha\;S\;W_\al W^\al
\eeq
where $S$ is a chiral multiplet whose $F$ component breaks supersymmetry.
$M$ is
the scale of the physics that mediates the breaking to the low energy sector.
As an example [9] one could imagine heavy (mass $M$) superfields charged under
the standard model gauge  group $SU_C(3)\times SU_L(2)\times
U_Y(1)$ that couple directly to the supersymmetry breaking sector.  The low
energy sector then feels the breaking of supersymmetry through usual gauge
interaction with the heavy sector.  In this case, $M$ can be substantially
lower
than the Planck scale.  Indeed, after supersymmetry breaking the dimension
5 operator leads to a gaugino majorana mass term
\beq
\frac{F_S}{M}\; \la^a\la^a
\eeq
where $a$ are gauge indices.  So that
\beq
m_\la\sim\frac{F_S}{M}\;.
\eeq
The scalar quarks,
leptons and higgs    masses appear through dimension 6 operators of the
form
\beq
\frac{1}{M^2}\;\int d^4\Tha\;S^*S~\,T^*T,
\eeq
 where the $T$ superfield stands
for the quarks, leptons and Higgs superfields.  After supersymmetry breaking, a
(mass)$^2$ term is generated of the form
\beq
\frac{|F_S|^2}{M^2}~~A_T^*\,A_T\;.
\eeq
As stated before we will assume that
\beq
\frac{ |F_S|}{M} \sim \cO \left(G_F^{-1/2}\right)\;.
\eeq
It is therefore possible to have $F_S$ as small as a TeV and $M$ of order 10
TeV.  This would then lead to a tremendously small mass $m_{3/2}$ for the
gravitino, $m_{3/2} \sim \cO\left(10^{-4}{\rm eV}\right)$.  Indeed the
gravitino
mass due to the mixing with the would be goldstino is
\beq
 m_{3/2}\sim\frac{|F_S|}{M_p}\,.
\eeq
Then by using expression (6),  $m_{3/2}$ can be reexpressed as
\beq
m_{3/2}\sim \frac{M}{M_p}\;G_F^{-1/2} .
\eeq
  Now in order for no fine tuning of
scalar Higgs masses, $m_{3/2}$ should not be heavier than a TeV.  In order to
discuss the cosmological consequences of gravitinos we need to consider their
interactions with other fields.    As is well known [10], the
lifetime
$\tau_{3/2}$ for a gravitino of mass $m_{3/2}$ larger than the typical
splitting $\Del M$ in supermultiplets, is
\beq
\tau_{3/2}\sim M_p^2/m^3_{3/2}\;.
\eeq
The lifetime $\tau_{3/2}$ then ranges from days to $10^4$ years for $m_{3/2}$
in
the range \break 10GeV $\ltsim m_{3/2}\ltsim$ 1TeV.  This presumes the lightest
superpartner to be {\it not} heavier than 10GeV.  As was pointed out in
previous studies [1-6, 10], this could be potentially fatal for nucleosynthesis
if gravitinos had earlier been in thermal  equilibrium.  Gravitinos lighter
than
the lightest superpartner could in principle decay to a neutrino and photon
through higher dimensional operators that violate $R$ parity and lepton number.
An example is the dimension 6 operator
\beq
 \frac{1}{M_p^2}\, \Psi^{\mu*} \bar\si^\nu\,
    L^a (F_{\mn})_{ab}\, H^b_u
\eeq
where $\Psi^\mu$ is the spin 3/2 component of the gravitino, $L^a$ is a
lepton doublet, $H_u^b$ is the Higgs doublet that gives mass to the up
quarks and $F_{\mu\nu\,b}{}^{\!\!\!a}$ is the $SU(2)$ field strength.
This operator
leads to a lifetime $\tau_{3/2}$ for the gravitino
\beq
\tau_{3/2}\sim\frac{M_P^4}{m_{3/2}^5}
\eeq
 which is much larger than the age of
the universe for
 $ m_{3/2}$ of $\cO\left(G_F^{-1/2}\right)$.
 Therefore, such gravitinos are stable and
the danger [1-6, 10], in this case, is that their present
energy density, $\rho_{3/2}$ (today), could be larger than the critical energy
density $\rho_c$.

The crucial information needed to discuss the cosmologial
implications of gravitinos, is their abundance.  In order to estimate this
abundance, we will need to calculate their interaction rate with the early
universe plasma. This is precisely where our analysis differs with
earlier estimates [2-6].  When the energy $E$ and momentum $|\bar{k}|$ of the
gravitino is larger than the typical splitting $\Del M$  in a multiplet, but
smaller than $\sqrt{F_S}$, the dominant interaction of the gravitino is through
its longitudinal component, the ``goldstino'' $\Psi_S^\al$.  In this range of
energies one can derive the interaction of $\Psi_S^\al$ by using a non linear
realization of supersymmetry [11].  Indeed, under a rigid supertransformation,
the goldstino transforms as:
\beq
\del\Psi_S^\al=\varepsilon^\al\lag F_S\rag + \ldots
\eeq
so that it must decouple at zero momentum, as do
goldstone bosons.  The leading terms in the interaction Lagrangians are then:
\beqra
L_{int}&=& \frac{1}{F_S}\,\left[\frac{1}{4}\,
\pa_\mu\Psi_S \si_{[\al,}\bar\si_{\beta]}\si_\mu\la^{*a}F^a_{\ab}\right.
\nonumber \\ && + (D_\rho A)^*\pa_\mu\Psi_S\si_\rho\bar\si^\mu\Psi_S-
\frac{ig}{2}\;(\pa_\mu \Psi_S)\,\si_\mu\,\la^{*} (A^*T_aA) \nonumber \\
&& \left.-2i(\pa\mu\Psi_S)\si^\mu\Psi_i^* \, \frac{\pa W}{\pa F^*_i}
\right] + h.c.
\eeqra
where $\la_a$ is a gaugino field, $F_{\mn}^a$ is the field strength and $A$ and
$\Psi$ are respectively scalar and fermion components of chiral multiplets.
One
can then calculate for example in this regime the cross section $\si$ for
$ {\rm gluon} + {\rm gluon} \rta {\rm gravitino} + {\rm gluino}$, see fig. 1.
As Fayet [12] pointed out, this cross section $\si$ is given by
\beq
\si \sim \al_S\; \frac{(\Del M)^2}{F_S^2}\,.
\eeq
Note that in the regime of energy $\Del M<E<\sqrt{F}$, one could
naively expect that due to the derivative coupling of the goldstino, the cross
section depends quadratically on the energy:
\beq
\si\sim\al_S\;\frac{E^2}{F_S^2} \;.
\eeq
This however does not occur, instead  $\si$ saturates to the behavior found in
expression (14).  The reason for this, is that in this regime, the
splitting $\Del M$ becomes negligible compared to the energy scale so that
supersymmetry is restored and the goldstino hence decouples.  This implies that
the leading energy dependence of the separate diagrams in fig. 1
cancels, leaving over (14).

In the cosmological context however, we are interested in the cross section at
high temperatures $T$ in the range $\Del M<T<\sqrt{F}$.  Unlike the high
energy,
zero temperature case, the cross section $\si$ in this thermal environment does
not saturate but instead grows as
\beq
\si\sim \al_S^3\; \frac{T^2}{F_S^2}\;.
\eeq
  The reason for the growth in the cross
section is that at finite temperature the bosonic states with momenta
$|\bar{k}|\ltsim  T$ have large occupations, whereas the fermionic states are
at
most occupied by one quantum.  At the leading order, depicted in fig. 1,
because of energy-momentum conservation the cancellation still occurs [16].
However, the next order in $\al_S$ allows for intermediate bosonic states
occupied by more than one quantum.  Therefore, even though the temperature
$T$ is larger than the splitting $\Del M$, the leading temperature behavior
at this next to leading order in $\al_S$, will not
cancel among the various diagrams.
This is because the different statistics of bosons and fermions ``break''
supersymmetry in a thermal environment.  The rate
$\Gam$ for such interactions in this temperature regime by dimensional
arguments
is
\beq
\Gam\sim\al_S^3\;\frac{T^5}{F_S^2}\,.
\eeq
The gravitino can therefore be in equilibrium at temperature
\beq
T\gtsim \frac{1}{\al_S}\,\left(\frac{F_S^2}{M_p}\right)^{1/3} \sim
\frac{1}{\al_S}\,m_{3/2}^{2/3}\,M_p^{1/3}\,.
\eeq
If the temperature of the universe did reach such value, then the number
density
of gravitino $n_{3/2}(T)\sim T^3$.  The energy density $\rho_{3/2}(T)$ would
then satisfy
\beq
\rho_{3/2}(T)\sim g^*(T)\,m_{3/2} T^3
\eeq
where $g^*$ is a dilution factor due to previous annihilations or decays, which
implies for stable gravitinos $\rho_{3/2}$ (today) $\sim
10^{-2}m_{3/2}(3\;10^{-4}\;{\rm eV})^3$.  We then recover     the
earlier literature result that gravitinos with masses $m_{3/2}< 1\,$keV satisfy
the constraint that their present energy density is bounded by $\rho_c$.  For
masses $m_{3/2}$ in the range 1keV $<m_{3/2}< 10$GeV, we find that these should
never have been in equilibrium.  Therefore, a simple upper bound on the maximum
Temperature $T_{max}$ is the Temperature $T_{eq}$ at which gravitino
equilibrates,
\beq
T_{eq}\sim \frac{1}{\al_S}\, m_{3/2}^{2/3} \; M_p^{1/3}\,.
\eeq
There are presumably large numerical factors in (17), typical of thermal
calculations involving large numbers of components, like in the Debye mass for
gluons [17].  This is why the factor of $\frac{1}{\al_S}\,$ does not
numerically affect much the bounds on $T_{max}$.

One can get a more serious bound by strictly implementing
$\rho_{3/2}$ (today) $<\rho_c$.  This is done by integrating the relevant
Boltzman equations. The result for
 $T_{max}$ as a function of $m_{3/2}  $ is depicted on the diagram
in fig. 2.  Note that for \break
$m_{3/2} \sim 10\,$ GeV the maximum temperature $T_{max}\sim \cO\, (10^5$ GeV).

For unstable gravitinos with masses in the range \break 10GeV $\ltsim
M_{3/2}\ltsim 1$ TeV the bound on $T_{max}$ is obtained by requiring that
nucleosysthesis not be affected.  This entails that during this period, the
gravitino doesn't dominate the energy density of the universe.
Also, their decays after nucleosynthesis should not generate more
entropy  than was already present, such as not to dilute the abundance of light
nuclei.  The latter bound is stronger in constraining the abundance of
gravitinos.  One should also consider the destruction of deuterium, and other
light nuclei by decay products of gravitinos [2, 3, 5, 6]. This was shown in
the
past, to lead to the most stringent bound on the ultimate temperature of the
universe.

The ratio of entropy before decay $S_i$, to after decay $S_f$ can be
estimated to
be
\beq
\frac{S_f}{S_i}\sim \frac{[n_{3/2}(\tau_{3/2}) \,
m_{3/2}]^{3/4}}{n_{3/2}\;(\tau_{3/2}) }\,.
\eeq
By equating the inverse lifetime of gravitinos with the expansion rate of the
universe dominated by these gravitinos one obtains
\beq
n_{3/2}\, (\tau_{3/2}) \sim \frac{m_{3/2}^5}{M_p^2}\,.\eeq
This leads to
\beq
\frac{S_f}{S_i}\sim\left(\frac{M_p}{m_{3/2}}\right)^{1/2}\,.
\eeq
Requiring that the decaying gravitinos not modify in any significant way the
already present entropy, leads to a bound on
$T_{max}$ as a function of the gravitino mass $m_{3/2}$ as depicted on the
diagram of fig. 3.  For a range of gravitino mass, 10GeV $\ltsim m_{3/2} \ltsim
1$ TeV, the ultimate temperature $T_{max}$ range is then
\beq
10^5 {\rm GeV} \ltsim T_{max} \ltsim 10^7 \;{\rm GeV}\;.
\eeq
If one considers instead, the bound from the destruction of light nuclei,
following previous authors [2, 3, 5, 6]  the result is even more dramatic
in that it predicts a maximal temperature $T_{max}$ in the  range
\beq
10^4 {\rm GeV} \ltsim T_{max} \ltsim 10^5 \;{\rm GeV}\;,
\eeq
for the same range of gravitino mass, as can also be seen in fig. 3.   These
results are obtained by rescaling the bounds appearing in earlier papers [3,6],
with the new temperature dependence of the gravitino abundance.

In conclusion the possible existence of gravitinos with masses \break
$m_{3/2}>1$
keV require low temperature baryogenesis.  Such late generation of a baryon
asymmetry was discussed earlier in a beautiful paper by Affleck and Dine [13].
Also, more recently, a flurry of activity around weak interaction baryogenesis
[14] provides another way to reconcile observation with the possible existence
of gravitinos with masses of order or smaller than  a TeV.  Note in addition,
that if the universe had experienced a period of inflation with cosmological
constant
$\La$,  the reheating temperature $T_R$ should then be smaller than $T_{max}$.
If it appears that the microwave anisotropy has a tensor component, then the
cosmological constant $\La$ is of $O[(10^{17}\;{\rm GeV})^4]$ [15].  This would
then imply an extremely inefficient reheating with
\beq
\frac{T_R}{\La^{1/4}}<\frac{T_{max}}{10^{17}\,{\rm GeV}}\,.
\eeq
Therefore the ratio ~$\displaystyle{\frac{T_R}{\La^{1/4}}}$~ should not exceed
an
extremely small number in the vicinity of $10^{-12}, ~10^{-13}$.  This in turn
requires an extremely weak coupling between the sector responsible for
inflation
and the standard model degrees of freedom.  This estimate is conservative since
the production of gravitinos by the inflationary sector during the exit of
inflation was neglected.

To avoid any bound on
the temperature of the universe, gravitinos should be lighter than 1 keV.  This
in turn implies low energy supersymmetry breaking at a scale $F_S$ comparable
or
smaller than (10$^6$GeV)$^2$. In this case, supersymmetry has to be mediated to
the low energy sector by forces other than gravity.

\section*{Acknowledgements:}

\indent\indent I would like to thank Michael Dine, Vadim Kaplunovsky, Andrei
Linde, Sonia Paban,
and Lenny Susskind for helpful conversations.  I would also like to thank Alex
Avram for pointing out ref. [12] and for useful conversations.

\section*{Reference}
\be
\item[{[1]}]  H. Pagels, T.R. Primack, \prl {48} {82}, {223}.

\item[{[2]}]  M.Y. Khlopov, A.D. Linde, \plb {138} {84}, {265}.

\item[{[3]}]  T. Ellis, J.E. Kim, D.V. Nanopoulos, \plb {145} {84}, {181}.

\item[{[4]}]  T Moroi, H. Murayama, M. Yamaguchi, \plb {303} {93}, {289}.

\item[{[5]}]  V.S. Berezinski, \plb {261} {91}, {71}.

\item[{[6]}]  M. Kawasaki, T. Moroi, HEP-PH -9403364, preprint.

\item[{[7]}] For a review, see H.P. Nilles, ``Supersymmetry, Supergravity and
Particle Physics," {\it Phys. Reports} {\bf 110} (1984), 1-162 and references
therein.

\item[{[8]}]  See for example P. Fayet, T. Iliopoulos, \plb {51} {74}, {461};\\
B. deWitt, D.Z. Freedman, \prl {35}  {75}, {827}; \\
E. Cremmer, S. Ferrara, L. Girardello, \npb {212} {83}, {413}.

\item[{[9]}]  M. Dine, W. Fischler, \npb {204} {82}, {346}; \\
J. Polchinski, L. Susskind, \prd {26} {82}, 45; \\
L. Alvarez-Gaum\'e, M. Claudson, M. Wise, \\\npb {207} {82}, {16}. \\
Note that in these models the squarks are naturally degenerate $\cO(\al)$.

\item[{[10]}] S. Weinberg, \prl {48} {82}, {1303}; \\
L.M. Krauss, \npb {227} {83}, {556}.

\item[{[11]}]  V.I. Ogievetski, Lectures at 10th Karpacz Winter School; \\
D.V. Volkov, V.D. Akulov, JETP Lett. {\bf 16} (1972), 438.

\item[{[12]}]  P. Fayet, \plb {84} {79}, {421}.

\item[{[13]}]  I. Affleck, M. Dine, \npb {249} {85}, {361}.

\item[{[14]}] A.G. Cohen, C.B. Kaplan and A.E. Nelson, \\{\it Annu. Rev. Nucl.
Part. Sci.} {\bf 43} (1993) 27-70; \\
M. Dine, ``Electroweak Baryogenesis: A Status Report'', SCIPP 94/03. To appear
in the ``Proceedings of the Third KEK Topical Conference on CP Violation, its
Implications to Particle Physics and Cosmology,'' 16-18 Nov. (1993) and
references therein.

\item[{[15]}]  V.A. Rubakov, M. Sazhin, A. Veryashin, \plb {115} {82},;\\
R. Fabbri, M. Pollock, \plb {125} {83}, {445};\\
B. Allen, \prd {37} {88}, {2078};\\
L. Abbott, M. Wise, \npb {224} {84}, {541}.

\item[{[16]}]  A.D. Linde and L. Susskind, private communication.

\item[{[17]}] H.A. Weldon, \prd {26} {82} {1394}.

\ee

\pagebreak

\nd
\section*{Figure Captions}

\begin{itemize}

\item[{\bf Fig. 1:}]  Feyman diagram is involved in the calculation for
the cross
section of\break  gluon $+$ gluon $\rta$ gluino $+$ ``goldstino''.

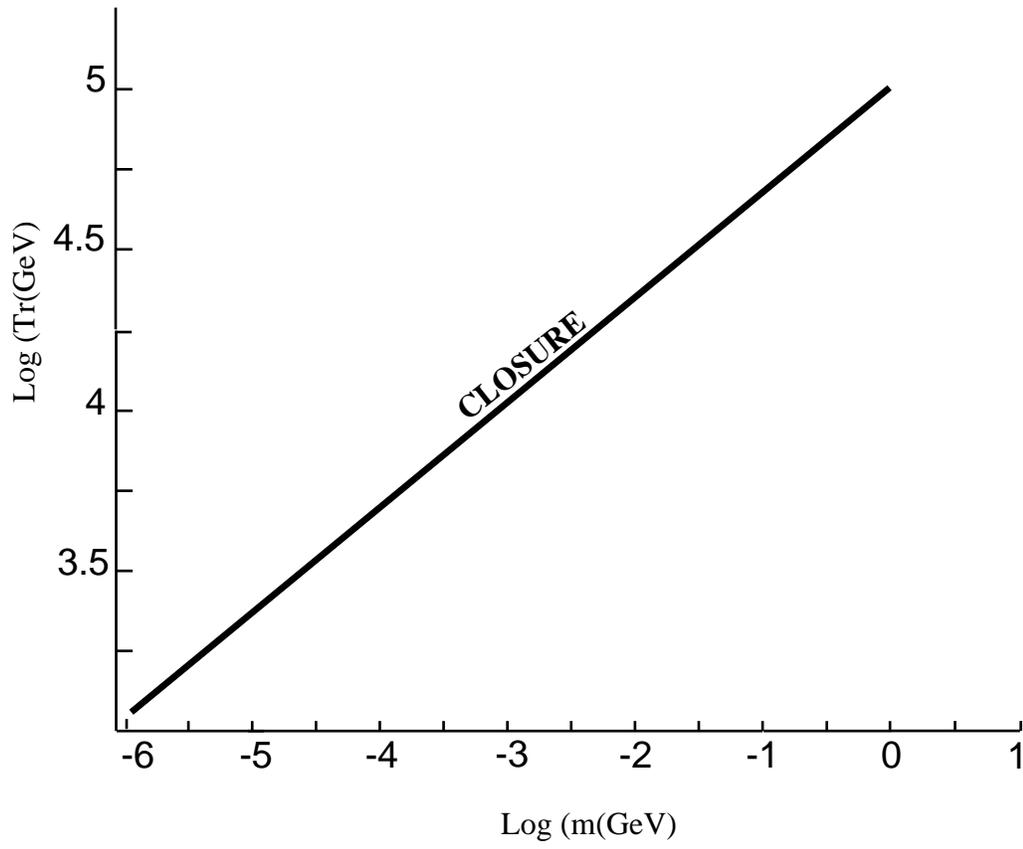
\item[{\bf Fig. 2:}] The diagram for the ultimate temperature $T_{max}$ as a
function of $m_{3/2}$, in the case of stable gravitinos.

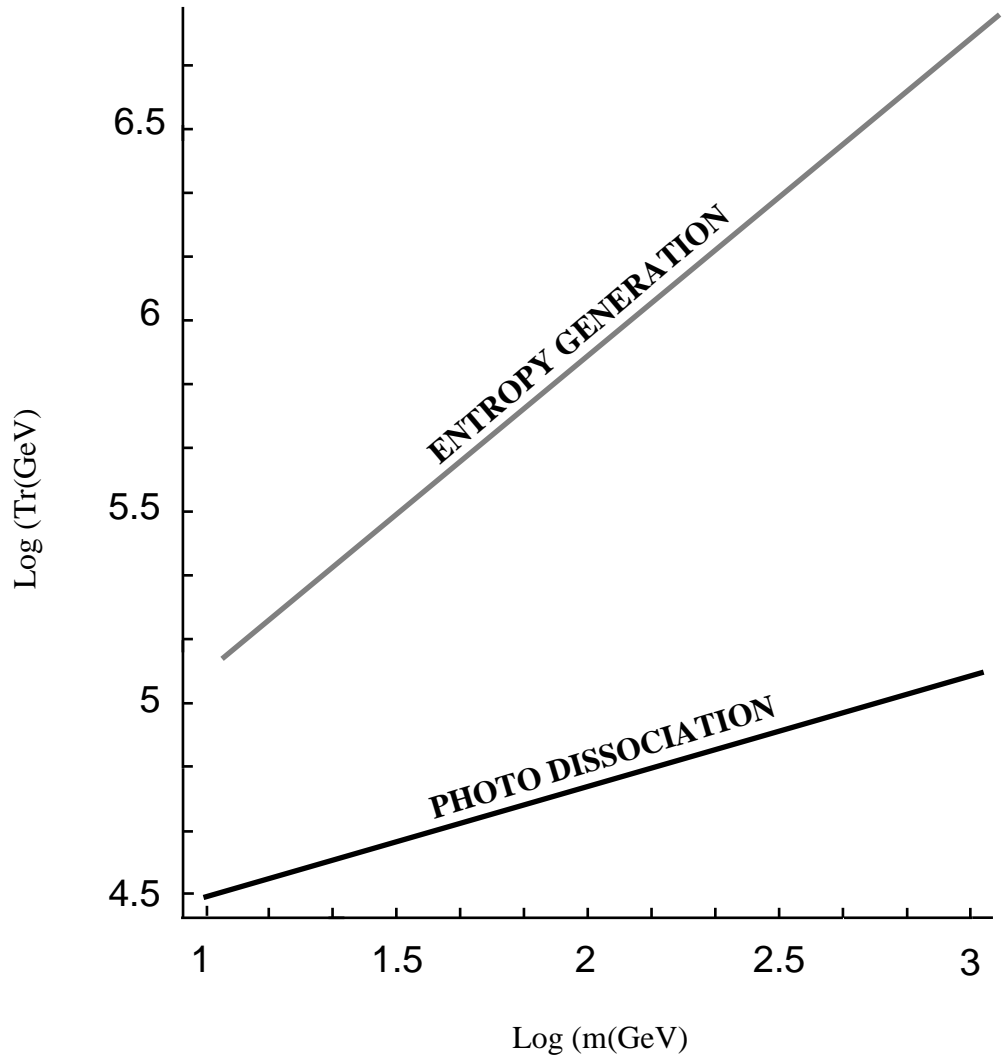
\item[{\bf Fig. 3:}]  The diagram for the ultimate temperature $T_{max}$ as a
function of $m_{3/2}$ in the case of unstable gravitinos.

\end{itemize}
\end{document}